\documentclass[%
reprint,
superscriptaddress,
 amsmath,amssymb,
prb,
]{revtex4-2}

\usepackage{graphicx}
\usepackage{dcolumn}
\usepackage{bm}
\usepackage{pifont}
\usepackage[version=3]{mhchem} 
\usepackage{braket}
\usepackage{float}
\usepackage{xcolor}
\usepackage{mathtools}
\usepackage{tabularx}
\DeclareMathSymbol{*}{\mathbin}{symbols}{"03} 
\DeclareMathSymbol{\ast}{\mathbin}{symbols}{"03}
\usepackage{hyperref}
\hypersetup{
    colorlinks=true,
    linkcolor=blue,
    filecolor=blue,      
    citecolor=blue,
    urlcolor=blue,
    pdftitle={Overleaf Example},
    pdfpagemode=FullScreen,
    }

\begin{document}

\preprint{APS/123-QED}
\title{A unified bonding  entropy model  to determine magnetic properties in graphene nanoflakes}

\author{Chang-Chun He}
\email{scuthecc@scut.edu.cn}
\affiliation{School of Physics and Optoelectronics, South China University of Technology, Guangzhou 510640, China}
\author{Jiarui Zeng}
\affiliation{School of Physics and Optoelectronic Engineering, Hainan University, Haikou 570228, China}
\author{Yu-Jun Zhao}
\affiliation{School of Physics and Optoelectronics, South China University of Technology, Guangzhou 510640, China}
\author{Xiao-Bao Yang}
\affiliation{School of Physics and Optoelectronics, South China University of Technology, Guangzhou 510640, China}

\date{\today}

\begin{abstract}
Graphene nanoflakes (GNFs) exhibit rich magnetic behaviors arising from two primary mechanisms: geometry frustration in non-Kekulé structures and electron delocalization-driven aromatic stabilization in Kekulé-type systems. Herein, we develop a unified bonding entropy model (BEM) to quantitatively characterize the magnetic properties in GNFs within a statistical framework, providing an entropy-based criterion for understanding and predicting  bond occupancy numbers and unpaired electron distributions. While non-Kekulé systems naturally favor high-spin configurations due to topological frustration, the BEM reveals that even Kekulé-type GNFs can exhibit magnetic character when the entropy gain from unpaired electrons outweighs the loss of aromatic stabilization. The model predictions show excellent agreement with density functional theory calculations in terms of spin density distributions and unpaired electron counts. Our results establish bonding entropy as a general guiding principle for designing carbon-based magentic materials with tunable magnetic properties.

\end{abstract}

\maketitle

\section{Introduction}

Graphene nanoflakes (GNFs), finite-sized fragments of graphene with well-defined edge geometries and symmetries, have emerged as a fertile platform for exploring unconventional magnetism in low-dimensional carbon-based systems \cite{PhysRevB.81.033403,PhysRevB.81.201402,PhysRevB.84.245403}. Unlike pristine graphene, which is a nonmagnetic semimetal due to its delocalized $\pi$-electron network and bipartite lattice symmetry \cite{Geim2007}, GNFs can host localized magnetic moments as a result of quantum confinement \cite{doi:10.1021/nl072548a}, edge effects \cite{PhysRevB.77.193410,PhysRevB.104.014404}, and topological frustration \cite{PhysRevLett.102.157201}. One of the prototypical magnetic GNFs is triangulene, the smallest triplet-ground-state polybenzenoid hydrocarbon with a triangular shape and zigzag edges, which was  synthesized in 2017  for the first time \cite{Pavliek2017}. Triangulene is characterized by a non-Kekulé structure, where the number of carbon atoms in the two sublattices of the graphene honeycomb lattice is imbalanced, leading to a net magnetic moment as predicted by Ovchinnikov’s rule \cite{Ovchinnikov1978} and Lieb’s theorem \cite{PhysRevLett.62.1201}. A series of experiments have confirmed that triangulenes with various sizes exhibit high-spin open-shell ground states \cite{Pavliek2017,doi:10.1021/jacs.9b05319,doi:10.1126/sciadv.aav7717}, exhibiting significant spin density localized along the edges.  Beyond triangulene, more complex GNF structures such as Clar’s goblet \cite{CLAR19726041}, composed of two fused triangulene units,  exhibit intrinsic magnetism in recent experiments \cite{Mishra2020,PhysRevLett.132.046201}. Despite their promising properties, the synthesis of such GNFs remains experimentally challenging and has so far been achieved  via the ultra-high vacuum on-surface synthetic techniques on noble metal substrates.

Kekulé diradicals represent another class of GNFs with intrinsic magnetic properties \cite{doi:10.1021/acs.chemrev.3c00406}. Although they can formally adopt closed-shell configurations, the emergence of unpaired spins is often driven by the gain in Clar’s aromatic sextets, which stabilize the underlying $\pi$-conjugated network \cite{doi:10.1021/jacs.2c09637}. The competition between maximizing aromatic stabilization and preserving conventional bonding patterns can give rise to diradical ground states, as confirmed by recent experiments \cite{https://doi.org/10.1002/anie.201502657,https://doi.org/10.1002/anie.202102757}. This mechanism plays a crucial role in governing the electronic configuration and magnetism of Kekulé-type GNFs.  Magnetic Kekulé structures often possess small singlet-triplet gaps \cite{PhysRevApplied.21.034057,Weng2024}, which significantly enhance optoelectronic performance by enabling efficient triplet exciton harvesting and promoting intersystem crossing in thermally activated delayed fluorescence (TADF) materials \cite{doi:10.1126/sciadv.aao6910}.

In general, magnetic order in GNFs is originated from topological frustration and delocalized $\pi$ electrons. For non-Kekulé structures, the number of unpaired electrons in bipartite lattices with sublattice imbalance has been successfully predicted by theoretical approaches \cite{Ovchinnikov1978,PhysRevLett.62.1201,doi:10.1021/acs.nanolett.9b01773,doi:10.1021/acs.jpclett.3c00570}. However, these methods
face three major limitations: (i) they struggle to capture the spatial distribution of spin and electron density; (ii) they lack the ability to quantitatively explain the magnetic origin in Kekulé-type structures; and (iii) they provide no clear criterion for assessing the relative stability of magnetic GNFs. Density functional theory (DFT) combined with the Hubbard model has revealed magnetic ground states in graphene nanomeshes \cite{PhysRevB.84.214404}, olympicene radicals \cite{PhysRevB.108.115113},  two-dimensional triangulene crystals \cite{PhysRevResearch.5.043226} and hydrogenated \ce{C60} \cite{PhysRevLett.106.166402}.  Note that the Hubbard model may converge to different spin configurations depending on the initial guess, and the DFT validation is ultimately required to determine the magnetic ground state. In spite of quantitative accuracy, the total energies of magnetic systems by DFT calculations also depend on  the initial spin configurations. Therefore, there is  an urgent need for a concise, physically transparent model that can quantitatively describe unpaired electron distributions and magnetic order in GNFs.


For the close-shell systems, we have previously proposed a statistical model to determine the electron density distribution and Hamiltonian for carbon \cite{10.1063/5.0244219} and boron \cite{PhysRevB.111.085408} nanostructures. In this work, we extend the bonding entropy model (BEM) to quantitatively investigate the magnetic order in three representative types of GNFs: triangulenes, Clar’s goblet, and Kekulé radical systems. Assuming that all valence electrons are distributed among C-C bonds, C-H bonds, and unpaired electrons localized on individual carbon atoms, the optimal electron distribution is determined by maximizing the system’s bonding entropy. Through this unified BEM, we can quantitatively predict both the electron density distribution and local magnetic moments across all three GNF types without introducing additional fitting parameters. Notably, the predicted occupancy numbers exhibit a linear correlation with C-C bond lengths, and the total energy of each structure shows excellent agreement with its bonding entropy, indicating that our model can providing deeper and clearer insights into the magnetic properties of these  GNFs.

\section{RESULTS and DISCUSSION}

\begin{figure}
    \centering
    \includegraphics[width=\linewidth]{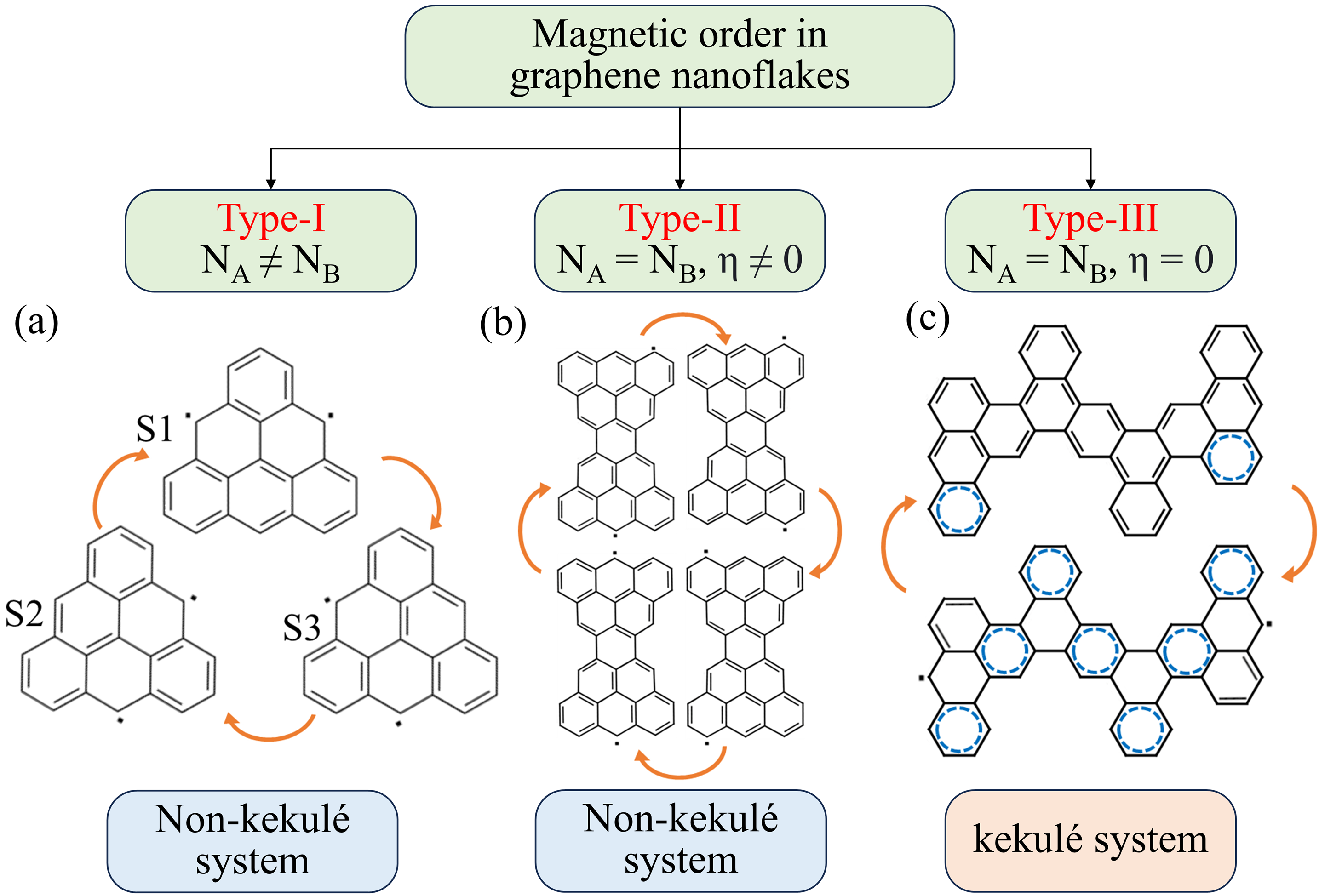}
    \caption{The classification of GNFs with magnetic order. (a) The three Kekulé resonance structures of the [3]triangulene molecule with two unpaired electrons. (b) The four Kekulé resonance structures of Clar’s goblet molecule with two unpaired electrons. (c) Schematic representation of the electronic configuration transition between the closed-shell and open-shell forms for the typical GNF, characteristic of a Kekulé diradical.}
    \label{fig1}
\end{figure}

To explore  the interatomic interactions in graphene nanoflakes, we begin by examining the [3]triangulene molecule,  where the number of carbon atoms in the two sublattices  is imbalanced, inducing a net magnetic moment $S=1$ with two unpaired electrons. As illustrated  Fig.~\ref{fig1}(a), the [3]triangulene possesses three distinct  resonance structures characterized by alternating single and double bonds, along with two unpaired electrons. These degenerate resonance  forms exhibit two-fold symmetry, violating the intrinsic three-fold symmetry of the [3]triangulene framework. Notably, due to the sublattice imbalance, not all carbon atoms meet the octet rule, yet two edge carbon atoms only possess seven electrons due to  the unpaired electrons, which are non-Kekul$\acute{\textrm{e}}$ systems. This type of GNFs are classified into type-I,  where the magnetic order is induced by sublattice imbalance and the magnetic moment is $S = |N_{A}-N_{B}|/2$.  

As shown in Fig.~\ref{fig1}(b), the type-II GNFs with magnetic order have balanced sublattice, but they can not be represented in Kekul$\acute{\textrm{e}}$ structures for two electrons are unpaired, inducing the magnetic order.  In the  view of  topology, \(\alpha\) and \(\beta\) denote the maximum numbers of nonadjacent vertices and edges, respectively, and the “nullity” is defined as \(\eta = |\alpha - \beta|\) \cite{PhysRevLett.102.157201}. When \(\eta = 0\) and \(\alpha = N/2\), all carbon atoms can be perfectly paired via nonadjacent bonds, known as ``perfect matching'', implying full pairing of all \(p_z\) orbitals. Nevertheless, \(\eta > 0\) indicates that some atomic sites remain unpaired in the optimal matching, which is referred to as ``topological frustration''. From the view of electron allocation, topological frustration  inevitably leaves some carbon atoms with only seven electrons as shown in Fig.~\ref{fig1}(b), and the unpaired electron delocalizes over the edge carbon atoms. These unpaired electrons induce local magnetic moments, leading to the emergence of magnetic ordering. 

While unpaired electron will contribute instability of a GNF structure,   an increased number of Clar’s sextets can enhance aromatic stabilization. The competition between these two effects may thus drive the emergence of magnetic ordering in GNFs. The third type of GNFs exemplifies the outcome of this competition, where magnetic ordering coexists with localized aromatic sextets, giving rise to intrinsic magnetic properties.  As shown in Fig.~\ref{fig1}(c), this type of GNFs with magnetic ordering have a series of Kekul$\acute{\textrm{e}}$ resonance structures, where all carbon atoms can satisfy the octet rule, but the  Clar’s $\pi$-sextets can  decouple paired electrons in a $\pi$ bond to unpaired states. This transformation can enlarger the localized aromaticity, where a greater number of Clar’s sextets corresponds to increased structural stability due to enhanced aromatic resonance.

Notably, magnetic GNFs inherently violate the octet rule, a unified bonding model is required to describe the three classes of magnetic GNFs, capable of determining a unique electron density distribution based on electron allocation principles. Although most carbon atoms in the GNFs approximately adhere to the octet rule, some carbon atoms possess only seven valence electrons due to the presence of unpaired spins. As a result, we should  go beyond the constraints of the conventional octet rule and incorporate a generalized bonding entropy model, where all valence electrons are allocated to chemical bonds and carbon atoms.

\subsection{The derivation of bonding  entropy model}

\begin{figure}
    \centering
    \includegraphics[width=\linewidth]{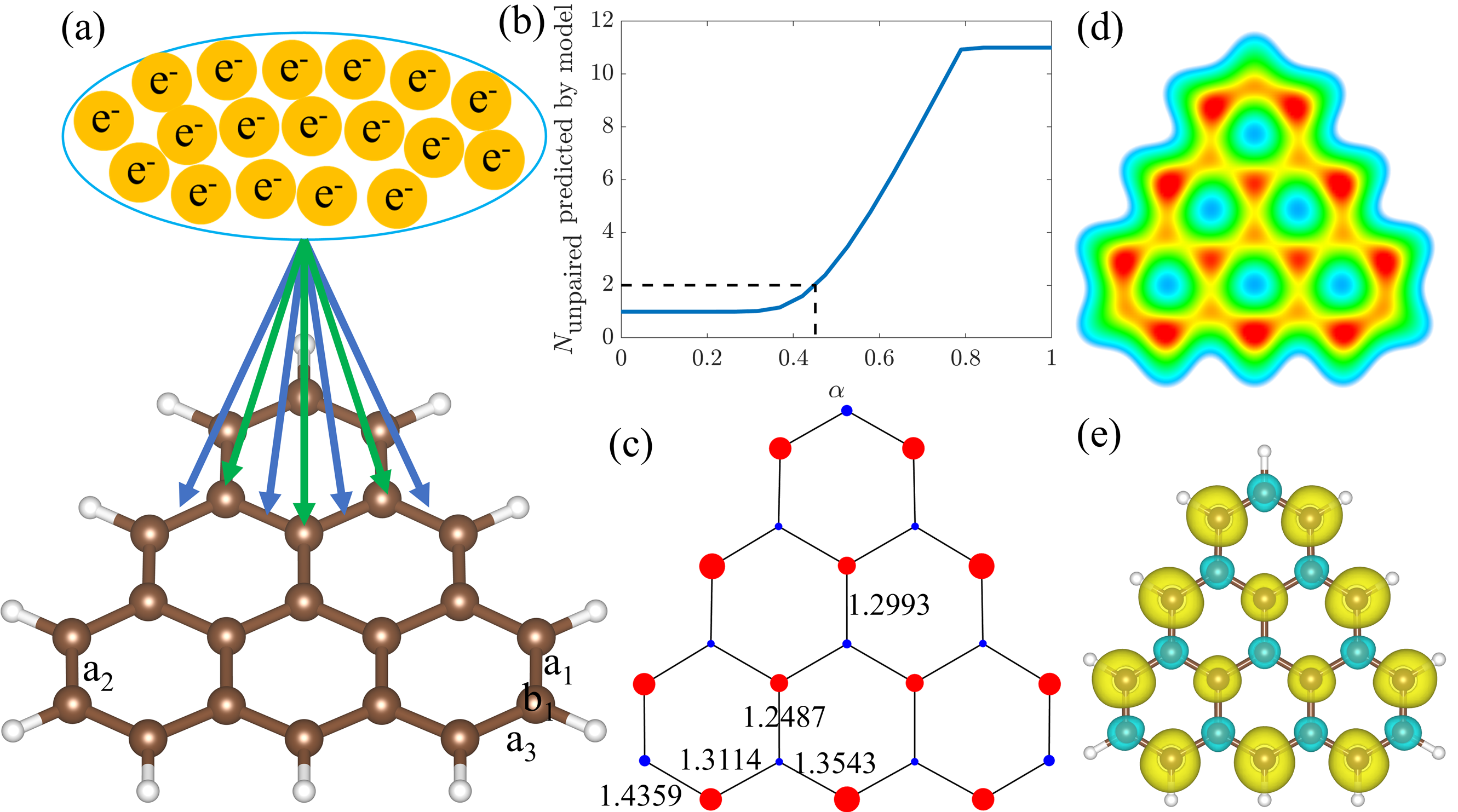}
    \caption{(a) Total electron allocation model for [3]triangulene molecule. (b) The $N_{\textrm{unpaired}}$  (c) The bonding free energy surface varied with $a_1$ and $a_2$ for [3]tringulene molecule. (d) The optimal occupancy number of each bond for [3]triangulene molecule, and the ON number is marked near the bond. The red and blue circles represents the number of unpaired electron in carbon atoms. The electron density (d) and spin density  (e)  distribution of [3]triangulene molecule calculated by DFT.}
    \label{fig2}
\end{figure}

Here, we assume that all electrons are delocalized and can move freely across all C-C bonds and carbon atoms. Within this framework, the number of electrons associated with each bond and each atom is treated as a variable. As illustrated in Fig.~\ref{fig2}(a), the total number of electrons in the system, \(N_{\textrm{ele}}\), is distributed over the C-C bonds and carbon atoms, where each bond is assigned \(n_i\) electrons (indicated by blue arrows) and each carbon atom carries \(b_i\) unpaired electrons (indicated by green arrows).   For each carbon atom, the four valence electrons are distributed among three neighboring C-C or C-H bonds and the carbon site  (as unpaired electrons), following the constraint $\frac{n_1}{2} + \frac{n_2}{2} + \frac{n_3}{2} + b_1 = 4$, where $n_1$, $n_2$, and $n_3$ represent the electron counts in the three nearest-neighbor chemical bonds, and $b_1$ denotes the number of unpaired electrons on the carbon atom. The factor of 1/2 accounts for the fact that each bond shares electrons equally between adjacent carbon atoms.  This constraint, referred to as the valence electron constraint, which  naturally conforms to the bonding rules of magnetic GNFs.

We define the electron density and spin density, \(p_i = n_i / N_{\textrm{ele}}\) and \(q_i = b_i / N_{\textrm{ele}}\), where \(n_i\) and \(b_i\) denote the number of electrons associated with the \(i\)-th bond and carbon atom, and \(N_{\textrm{ele}}\) is the total number of electrons. \(N_{\textrm{bond}}\) and \(N_{\textrm{C}}\) represent the total number of C-C bonds and carbon atoms in the system, respectively.  The total unpaired electrons $N_{\textrm{unpaired}}$ are defined by $\sum_i^{N_{\textrm{C}}}b_i$.  For a magnetic GNFs, there exists a series of bonding configurations (or electronic states) that satisfy the valence electron constraint, as illustrated in Fig.~\ref{fig1}. From a statistical standpoint, the most probable configuration is the one with the greatest number of indistinguishable microstates, corresponding to the maximum  entropy\cite{10.1063/5.0244219,PhysRevB.111.085408}. Here, we propose a bonding entropy model (BEM) to solve the electron density and spin density, where the bonding entropy is represented by:

\begin{equation}
    S = -\left[\sum_i^{N_{\textrm{bond}}} p_i\log(p_i)+\alpha\sum_i^{N_{\textrm{C}}}q_i\log(q_i)\right], 
    \label{Sb}
\end{equation}

Particularly, the contributions of bonding and unpaired electrons to the total bonding entropy should be assigned different weights, reflecting their distinct roles in the electronic structure.  Valence electrons are assumed to distribute themselves as uniformly as possible, corresponding to the maximization of the Shannon entropy over all bonds and atoms \cite{10.1063/5.0244219}. This uniform distribution promotes electron delocalization and enhances structural stability, reflecting the intrinsic tendency of the system to favor its most energetically favorable configuration. Accordingly, the optimal electron density distribution is obtained by maximizing the total bonding entropy of the system.

\begin{figure*}
    \centering
    \includegraphics[width=\linewidth]{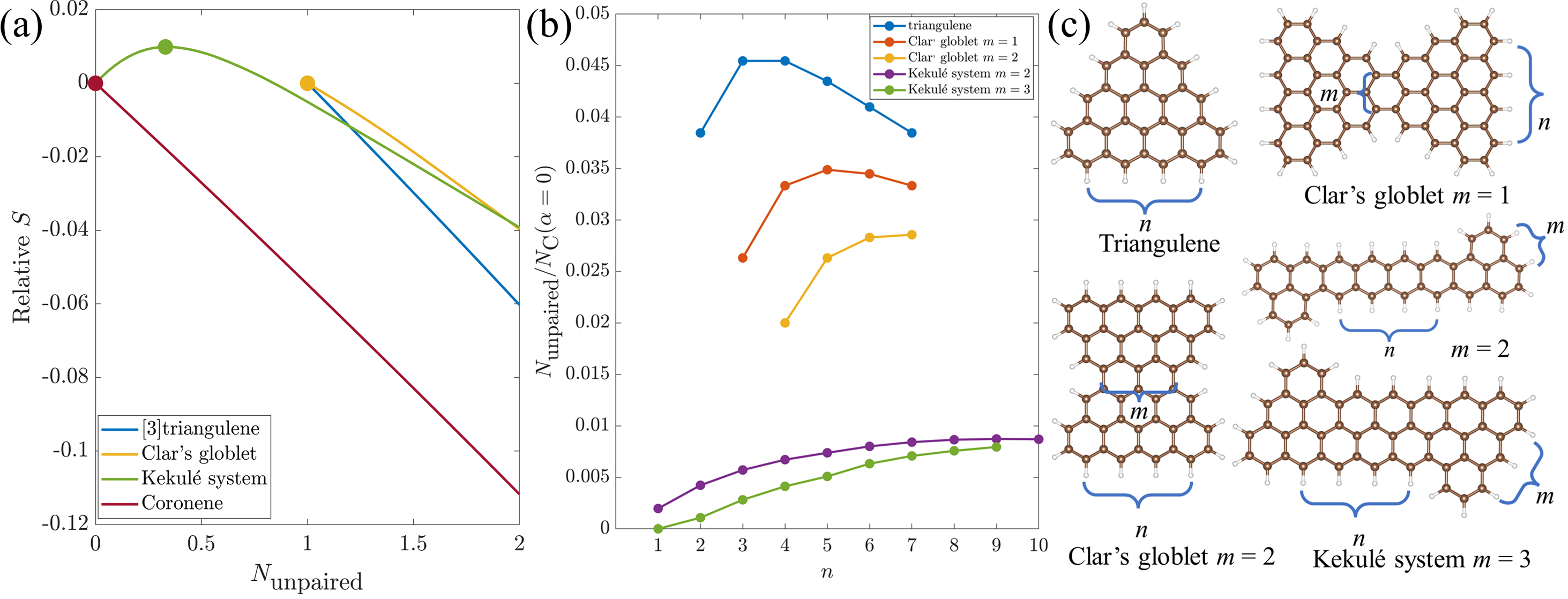}
    \caption{(a) Relationship between the number of unpaired electrons $N_{\textrm{unpaired}}$ and the relative bonding entropy $S$ for representative magnetic GNFs. (b) Scaled unpaired electron number $N_{\textrm{unpaired}}/N_{\textrm{C}}$ as a function of the structural parameter $n$ defined in (c), illustrating trends across different classes of magnetic GNFs. (c) Representative families of magnetic GNFs with tunable size parameter $n$.}
    \label{fig3}
\end{figure*}

Apparently, the parameter \(\alpha\) plays a crucial role in identifying the electron density distribution that maximizes the bonding entropy \(S\). Taking the [3]triangulene molecule as an example, when \(\alpha = 0\), the model predicts only one unpaired electron (see Fig.~\ref{fig2}(b)), which contradicts  that the [3]triangulene possesses two unpaired electrons. To recover the correct number of unpaired electrons, which originates from the sublattice imbalance, it is necessary to choose an appropriate value of \(\alpha\). The optimal value is found to be \(\alpha = 0.451\), which corresponds to exactly two unpaired electrons as shown in Fig.~\ref{fig2}(b). Particularly, it is important to emphasize that \(\alpha\) is not a freely adjustable parameter, which is determined by the intrinsic number of unpaired electrons in the system. For instance,  when \(\alpha > 0.8\), the BEM predicts 11 unpaired electrons, implying that  each carbon atom hosts an unpaired spin, which is clearly unphysical and inconsistent with known results.  This also highlights that the contribution weight of unpaired electrons to the bonding entropy must be significantly smaller than that of bonding electrons for a physically reasonable description.  Moreover, different systems require different values of $\alpha$, as the number of unpaired electrons varies, but remains uniquely determined by the system’s spin numbers.

With the optimal \(\alpha\), the model yields the accurate occupancy number for each C-C bond. Among the candidate solutions S1, S2, and S3 as shown in Fig.~\ref{fig1}(a), each exhibits a lower bonding entropy, while the global maximum entropy corresponds to the configuration illustrated in Fig.~\ref{fig2}(d). The occupancy numbers (ON), defined as \( n_i/2 \), are indicated near the C-C bonds, where ON = 1 corresponds to a single bond and ON = 2 to a double bond.  According to Ovchinnikov’s rule  \cite{Ovchinnikov1978}, sublattices A and B exhibit opposite spin orientation, with sublattice A being spin-up and sublattice B spin-down. This sublattice-resolved spin distribution is clearly captured in Fig.~\ref{fig2}(d), where neighboring atomic sites show opposite spin moments, visualized by red and blue circles whose radii represent the magnitude of unpaired electron density at each carbon site. The predicted electron density distribution \(p_i\) and the spatial distribution of unpaired electrons are in  agreement with DFT results, demonstrating that the BEM effectively captures the magnetic properties of the [3]triangulene molecule. Note that the optimal electron and spin density distributions predicted by BEM agree well with DFT results [Figs.~\ref{fig2}(d), (e)] as detailed in SM, demonstrating the predictive accuracy of BEM.  The implementation of this model is available  at    \href{https://github.com/ChangChunHe/max-entropy-for-magnetic-GNFs}{GitHub}.

The BEM can be understood as a model that maximizes the Shannon information entropy associated with electron distributions across C-C bonds and carbon atoms  \cite{10.1063/5.0244219}, where the coefficient $\alpha$ balances the weights of bonding and unpaired electron to the total bonding entropy. In \(\pi\)-conjugated GNF systems, electrons naturally favor a uniform distribution, and the  unpaired electrons of each carbon atom  corresponds to the number of locally unpaired electrons. This uniformity in distribution enhances the overall electron delocalization and decrease the electrostatic potential, thereby improving structural stability and enabling a faithful representation of the molecule’s intrinsic electronic and magnetic characteristics.

For GNFs, the presence of unpaired electrons inherently leads to magnetic ordering. Here, we investigate the relationship between the number of unpaired electrons, $N_{\textrm{unpaired}}$,  and $S (\alpha=0)$ across three representative types of magnetic GNFs in Fig.~\ref{fig3}(a). Notably, $N_{\textrm{unpaired}}$  of both [3]triangulene (Fig.~ \ref{fig1} (a)) and Clar's goblet (Fig.~ \ref{fig1} (b)) is greater than 1, and their bonding entropy decreases monotonically with increasing $N_{\textrm{unpaired}}$ due to the zero weight of unpaired electrons. Interestingly, for the type-III magnetic GNF, $S$ (at $\alpha=0$) first increases and then decreases with $N_{\textrm{unpaired}}$, suggesting that although unpaired electrons do not contribute directly to the entropy, their presence enables a more uniform distribution of bonding electrons, thereby enhancing the total bonding entropy and stabilizing local aromaticity. In contrast, for the non-magnetic coronene, $S$ (at $\alpha=0$) decreases monotonically with increasing $N_{\textrm{unpaired}}$, indicating that the emergence of magnetic character is energetically unfavorable. Therefore, we can classify the GNFs into different types according to the  bonding entropy at $\alpha=0$.
By evaluating $N_{\textrm{unpaired}}$ per carbon atom, magnetic GNFs can be effectively categorized into three types based on the BEM predictions at $\alpha = 0$, as illustrated in Fig.~\ref{fig3}(b). We use the structural order $n$ and $m$, defined in Fig.~\ref{fig3}(c), to describe the size of typical GNFs. With $n$ increases, type-I structures exhibit the largest values of $N_{\textrm{unpaired}}(\alpha = 0)/N_{\textrm{C}}$, followed by type-II, whereas type-III Kekulé-type systems possess the smallest values, which suggests that entropy-based metric thus offers a physically intuitive and quantitative criterion for classifying magnetic GNFs.

\subsection{The prediction of magnetic GNFs in non-Kekulé systems}
\begin{figure}
    \centering
    \includegraphics[width=\linewidth]{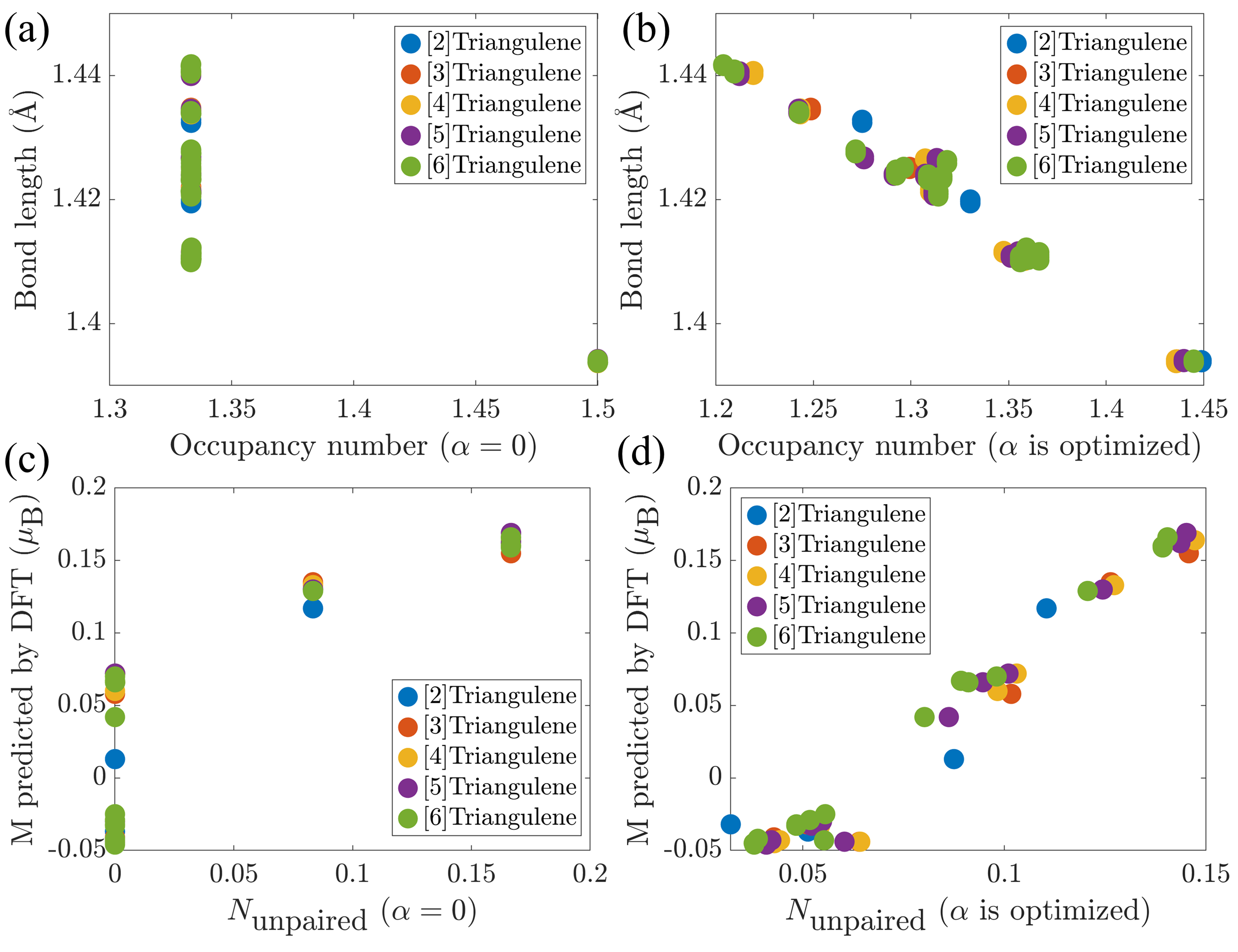}
    \caption{Relationship of ONs predicted by BEM with bond lengths predicted by DFT for triangulenes at (a) \(\alpha = 0\) and (b) optimized \(\alpha\). Relationship of unpaired electrons predicted by BEM with  magnetic moments predicted by DFT at (c) \(\alpha = 0\) and (d) optimized \(\alpha\).}
    \label{fig4}
\end{figure}

The above discussion shows that $\alpha$ is pivotal for accurately describing the interaction in [3]triangulene,  where the weight coefficient $\alpha$ is absolutely determined under the constraint of the number of unpaired electrons. In the following,  we will show the importance of $\alpha$ in predicting the bond length and total energy of GNFs with magnetic ordering. The occupancy numbers reflect the number of bonding electrons of C-C bonds, in which the larger ONs correspond to stronger interaction and shorter bond length. We take a series of triangulenes as the example to demonstrate the predictive ability of BEM. When $\alpha=0$, the ONs of all C-C bonds for triangulenes are 4/3, 1.5 as shown in Fig.~\ref{fig4}(a), which can not distinguish different C-C bonds, because $N_{\textrm{unpaired}}$ is not correct when $\alpha = 0$. The [$n$]trinagulene has $n-1$ unpaired electrons according the sublattice imblance. After $\alpha$ is optimized, it is clear that ONs have good linear relationship with C-C bond length.  Further, when $\alpha = 0$, the BEM also failed to predict the magnetic moment for each carbon atoms (see Fig.~\ref{fig4}(c)), after $\alpha$ is optimized, the predicted  magnetic moment for each carbon atoms in  triangulenes are in good agreement with the results from DFT as show in Fig.~\ref{fig4}(d). These  results  suggest that $\alpha$ is pivotal to enhance the predictive ability for BEM, in which the optimal  $\alpha$ corresponds to the true number of unpaired electrons.

\begin{figure*}
    \centering
    \includegraphics[width=\linewidth]{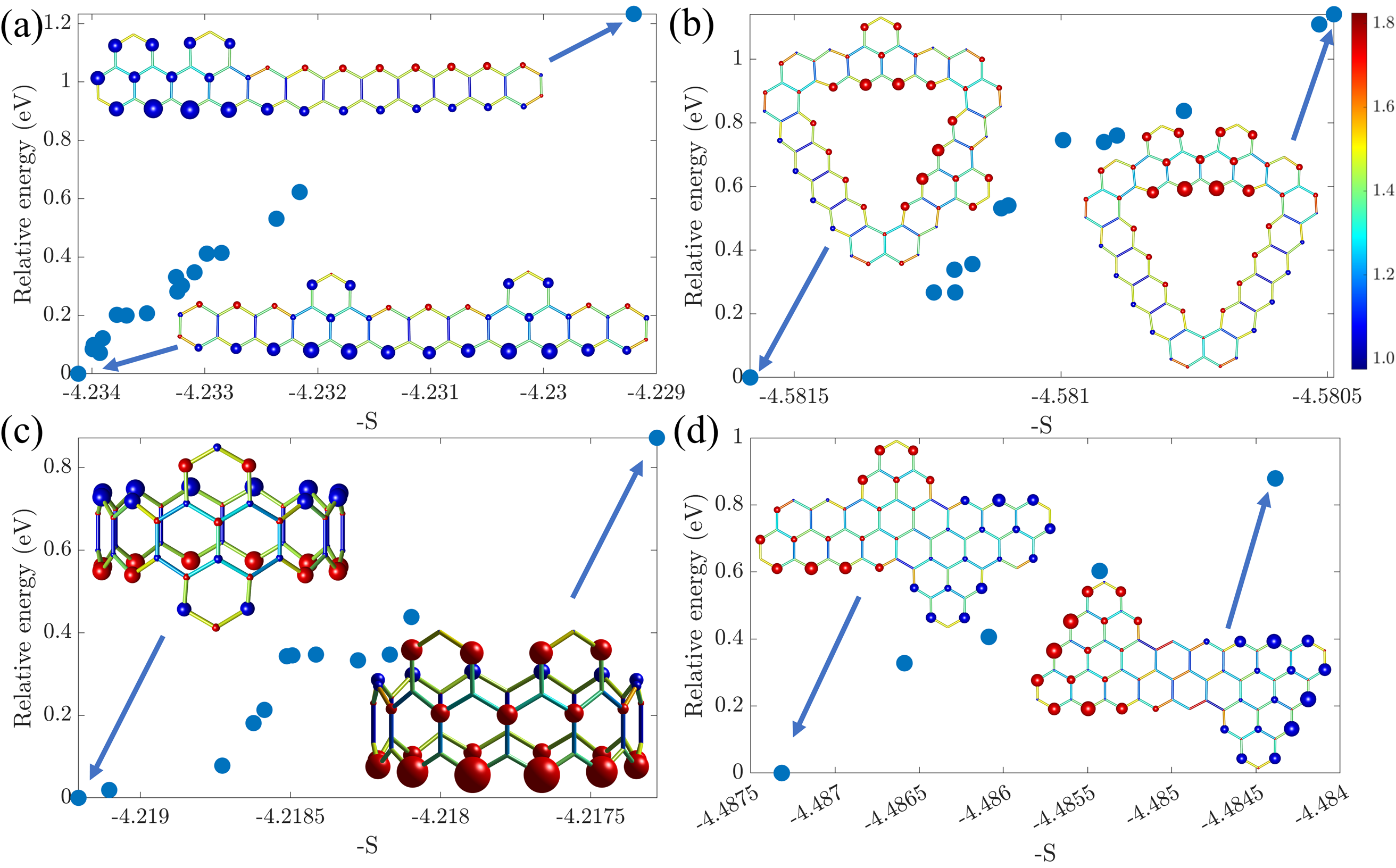}
    \caption{The relationship between relative  energy calculated by DFT  and bonding entropy for (a) a linear acene with two additional hexagonal rings; (b) a cycloarene with two additional hexagonal rings; (c) a carbon nanobelt with two additional hexagonal rings; (d) The magnetic GNFs with balanced sublattice.   The right colorbar corresponds the ONs of C-C bonds.}
    \label{fig5}
\end{figure*}

Since the BEM can accurately predict the bond length and local magnetic moments,  next we will show the maximal $S_{\textrm{b}}$ is capable to understand the structural stability of magnetic GNFs. We first constructed a linear acene consisting of 12 fused benzene rings, with two appended hexagonal rings. These isomers hold two unpaired electrons, where $\alpha$ can be optimized for all isomers.  As shown in Fig.~\ref{fig5}(a), the negative bonding entropy $-S$ linearly correlates with the total energy, indicating that the predictive power of bonding entropy. The most stable configuration is located at the lower right corner of Fig.~\ref{fig5}(a), where the electron density distribution is the most uniform, achieving maximal electron delocalization and corresponding to the largest bonding entropy. In contrast, the most unstable configuration occurs when the two additional hexagonal rings are positioned adjacent to each other, resulting in the lowest bonding entropy. This arrangement also induces the strongest steric repulsion between hydrogen atoms, further contributing to the instability. The distribution of magnetic moments is more uniform in the ground-state structure compared to the high-energy configurations, which partially mitigates repulsive interactions and enhances  structural stability. Similarly, we constructed the cycloarene and carbon nanobelt structures, with two additional hexagonal rings appended onto the main framework, which induce two unpaired electrons. After determining the appropriate \(\alpha\) corresponding to two unpaired electrons, we calculated the bonding entropy for each structure. The results show a strong correlation between the bonding entropy and the total energy of the structures based on  DFT. The most unstable structures in Figs.~\ref{fig5}(b) and (c) are both caused by the presence of two neighboring hexagonal rings, which lead to an uneven electron density distribution and significant steric repulsion between adjacent hydrogen atoms, indicating that the model reliably captures the relationship between electronic configuration and structural stability.

As shown in Figs. 5(a-c), the  types of GNFs with magnetic order exhibit sublattice imbalance. Next, we consider GNFs with balanced sublattices but magnetic ordering induced by topological frustration. To explore the relationship between bonding entropy and structural energy, we constructed a series of topologically frustrated GNFs by attaching one [2]triangulene unit to each side of the main framework (above and below), thereby preserving sublattice balance while introducing two unpaired electrons. As shown in Fig.~\ref{fig5}(d), the bonding entropy provides a reliable descriptor for the structural stability of these topologically frustrated graphene nanoflakes.  Therefore, the BEM can excellently predict the stability among widespread GNFs with magnetic order. The highest-energy structure features [2]triangulene units at the outermost edges, resembling two opposing [4]triangulene fragments. In this arrangement, spins of identical orientation are concentrated on the same side, amplifying repulsive interactions and thereby reducing structural stability. The ground-state structure allows for a more delocalized distribution of local magnetic moments and electron density, resulting in a more uniform bonding entropy across the system, which enhances structural stability.

\subsection{The prediction of magnetic GNFs in Kekulé systems}
We have discussed magnetic GNFs belonging to non-Kekulé systems, characterized by high-spin ground states. However, the synthetic challenges associated with these structures hinder their practical application in spintronics \cite{Pavliek2017,Mishra2020}. In contrast, Kekulé diradicals feature more stable radical characters, where the open-shell configuration appears  when the aromatic stabilization from additional Clar’s sextets outweighs the closed-shell preference \cite{doi:10.1021/jacs.2c09637}. Importantly, not all Kekulé systems exhibit radical character, such behavior emerges  when closed-shell configurations fail to provide sufficient local aromaticity, whereas the open-shell state enhances aromatic stabilization and overall stability.

\begin{figure}
    \centering
    \includegraphics[width=\linewidth]{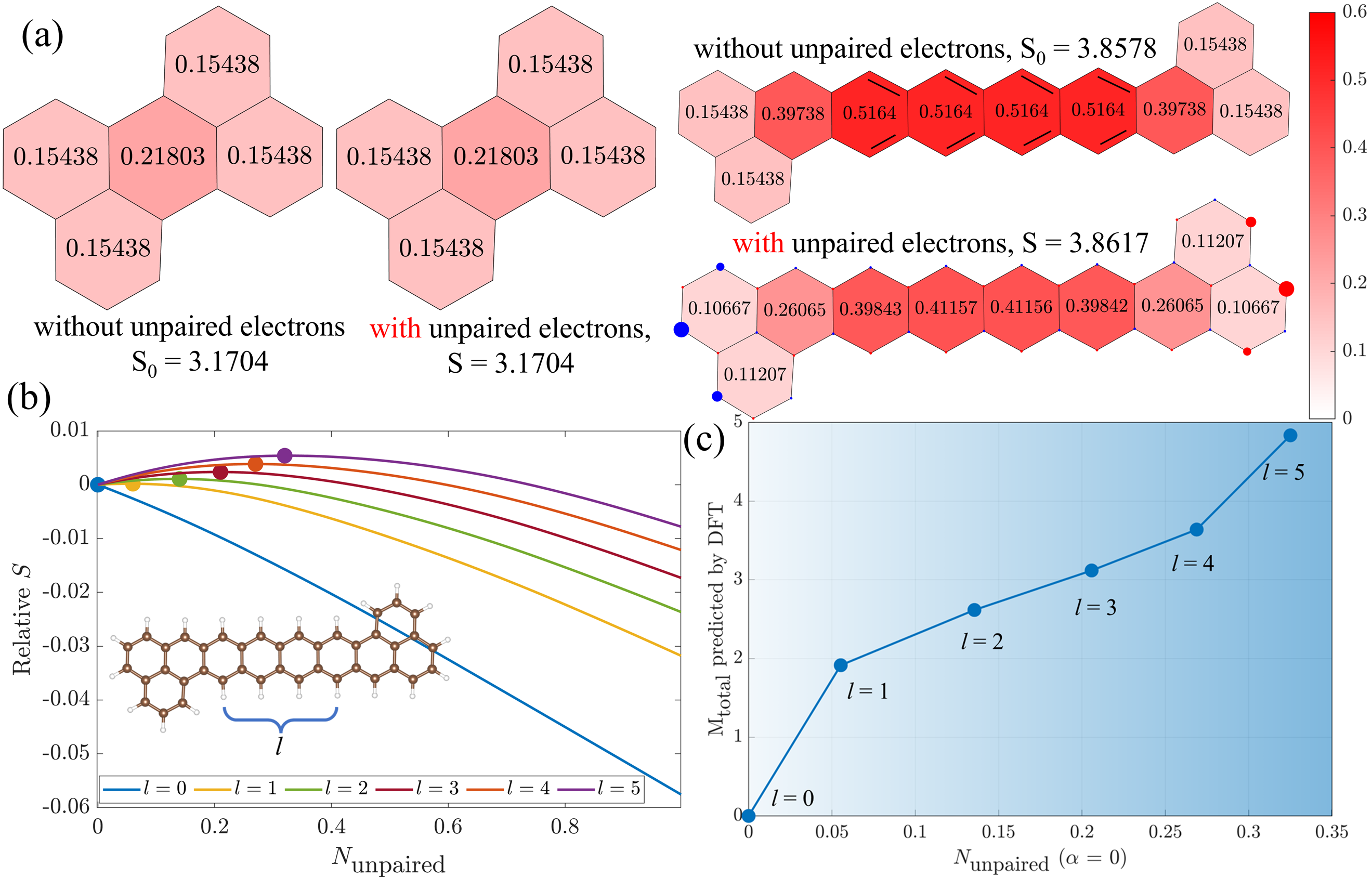}
    \caption{(a) Standard deviation of orbital occupation numbers (ONs) in two typical structures, calculated with (open-shell) and without (closed-shell) unpaired electrons in BEM. (b) Variation of relative bonding entropy $S$ and the number of unpaired electrons $N_{\textrm{unpaired}}$ for various GNFs with structural size parameter $l$. (c) Linear relationship between the magnetic moment obtained from DFT calculations and  $N_{\textrm{unpaired}}(\alpha = 0)$ predicted by BEM. }
    \label{fig6}
\end{figure}

The BEM offers an intuitive framework to capture this transformation and assess whether a Kekulé system favors a magnetic ground state. Specifically, comparing the bonding entropy $S_0$, with (defined by Eq.~\ref{Sb}) and without (detailed in section III of  SM) unpaired electron term, provides a clear criterion. For example, in perylene (left panel of Fig.~\ref{fig6}(a)), the equality $S_0 = S$ implies that adding the unpaired electron term yields no gain in bonding entropy, indicating the absence of unpaired electrons, consistent with DFT and experimental observations \cite{doi:10.1021/acs.chemrev.5b00188}. Furthermore, the small standard deviation of ONs across hexagonal rings supports the conclusion that no additional unpaired electrons are needed to enhance local aromaticity as shown in the left panel of Fig.~\ref{fig6}(a). 

We consider another molecule constructed by connecting two centrosymmetric [2]triangulene units via four acene bridges.  If  all valence electrons are confined to C-C and C-H bonds, the central hexagons display large ON deviations, signaling weak aromaticity and diminished stability and the bonding entropy of this molecule is 3.8578. The central hexagonal rings adopt a fixed alternating single-double bond pattern, rather than the delocalized resonance structure characteristic of benzene, thereby contributing little to the whole aromatic stabilization. In contrast, we can allow electron localization on carbon atoms, but the bonding entropy is not contributed by the unpaired electrons corresponding to $\alpha=0$ in BEM. This  leads to a significant reduction in ON fluctuations and larger bonding entropy of 3.8617,  enhancing aromatic stabilization as shown in Fig.~\ref{fig6}(a).  The comparison reveals a key physical principle: when fewer electrons are allocated to chemical bonds, the system can attain a higher bonding entropy, indicating that the emergence of unpaired electrons may not only promote a more uniform electron density distribution but also stabilize the structure as a result of the aromatic gain. 

To systematically investigate the evolution of radical character, we define $n$ as the number of hexagonal linkers bridging two centrosymmetric [2]triangulene units, with $n = 0$ corresponding to the perylene molecule as depicted in the inset of Fig.~\ref{fig6}(b). The reference entropy $S_0$, computed without the unpaired electron term, serves as a baseline of zero. By permitting a fraction of electrons to localize on carbon atoms as unpaired spins, we evaluate how the bonding entropy $S$ evolves with the number of unpaired electrons, in which only electrons in C-C bonds contribute to the bonding entropy. As shown in Fig.~\ref{fig6}(b), for $n = 0$, $S$ decreases monotonically as the number of unpaired electrons increases, suggesting a non-radical ground state. In contrast, for $n > 0$, $S$ exhibits a pronounced maximum at a specific number of unpaired electrons (highlighted by circles in Fig.~\ref{fig6}(b)). In fact, the maximum corresponds to the BEM with $\alpha=0$.  This behavior implies that, although fewer electrons contribute directly to bonding, the enhanced uniformity in electron density leads to greater bonding entropy. The system thus favors the emergence of unpaired electrons as a means of achieving increased structural stability.  In BEM, when $\alpha$ is zero, the optimal number of unpaired electrons  identifies the maximum of $S$, corresponding to the entropy maximum in Fig.~\ref{fig6}(b) as detailed in SM.  To show the predictive ability of BEM, we use a series of molecules to calculated the $N_{\textrm{unpaired}}$ when $\alpha=0$, which increases with \(n\), indicating an increasingly pronounced radical character. $N_{\textrm{unpaired}}$ linearly correlates with the total magnetic moment calculated by DFT as shown in Fig.~\ref{fig6}(c),  demonstrating the predictive ability of BEM.

\section{Conclusions}

In summary, we have developed a unified bonding entropy model (BEM) that quantitatively links the number of unpaired electrons to the magnetic stability of graphene nanoflakes (GNFs). By introducing a statistical framework constrained by the valence electron constraint, the BEM captures the interplay between electron pairing, bond occupancy number, and entropy maximization. Our results reveal that unpaired electrons may emerge not only from topological frustration in non-Kekulé systems but also from entropy gain in electron density distributions for Kekulé-type structures, in excellent agreement with density functional theory calculations in terms of spin density distributions and unpaired electron counts. Across a broad class of GNFs, bonding entropy emerges as a robust descriptor for structural and magnetic stability. Notably, magnetic GNFs can be effectively classified into two categories: non-Kekulé systems, where magnetism is driven by topological frustration, and Kekulé-type systems, where it originates from delocalization-induced aromatic stabilization. These results establish bonding entropy as a general guiding principle for designing carbon-based magnetic materials, paving the way for entropy-driven strategies in the discovery of magnetic systems and spintronic applications.

\begin{acknowledgments}
This work is supported by the Guangdong Basic and Applied Basic Research Foundation (Grants No. 2023A1515110894), the National Natural Science Foundation of China (Grant No. 12074126, No. 12474228), the Start-up Research Foundation of Hainan University (Grant No. XJ2500000571). This work is partially supported by High Performance Computing Platform of South China University of Technology. 
\end{acknowledgments}

%

\end{document}